\documentclass[twocolumn,superscriptaddress,showpacs]{revtex4}

\usepackage{graphicx}%
\usepackage{dcolumn}
\usepackage{amsmath}
\usepackage{amssymb}
\usepackage{bm}
\usepackage{color}

\begin{document}



\preprint{submitted to Physical Review X}
\title{
Three-dimensional electronic structures and the metal-insulator transition \\
in  Ruddlesden-Popper iridates
}

\author{A. Yamasaki}
\email[Author to whom correspondence should be addressed. \\
E-mail: ]{yamasaki@konan-u.ac.jp}
\affiliation{Faculty of Science and Engineering, Konan University, Kobe 658-8501, Japan }

\author{H. Fujiwara}
\author{S.~Tachibana}
\affiliation{Graduate School of Engineering Science, Osaka University, Toyonaka, Osaka 560-8531, Japan}

\author{D. Iwasaki}
\author{Y. Higashino}
\author{C. Yoshimi}
\author{K.~Nakagawa}
\affiliation{Graduate School of Natural Science, Konan University, Kobe 658-8501, Japan }

\author{Y.~Nakatani}
\author{K. Yamagami}
\author{H. Aratani}
\affiliation{Graduate School of Engineering Science, Osaka University, Toyonaka, Osaka 560-8531, Japan}

\author{ O. Kirilmaz}
\author{M.~Sing}
\author{ R. Claessen}
\affiliation{Physikalisches Institut and R\"ontgen Center for Complex Material Systems (RCCM), Universit\"at W\"urzburg,  D-97074 W\"urzburg, Germany}

\author{H.~Watanabe}
\affiliation{Waseda Institute for Advanced Study, Shinjuku, Tokyo 169-8050, Japan}

\author{T.~Shirakawa}
\author{S. Yunoki}
\affiliation{RIKEN Center for Emergent Matter Science (CEMS), Wako, Saitama 351-0198, Japan}

\author{A. Naitoh}
\author{K. Takase}
\affiliation{College of Science and Technology, Nihon University, Chiyoda, Tokyo 101-8308, Japan}

\author{J. Matsuno}
\affiliation{RIKEN Center for Emergent Matter Science (CEMS), Wako, Saitama 351-0198, Japan}

\author{H. Takagi}
\affiliation{Department of Physics, University of Tokyo, Tokyo 113-0033, Japan}
\affiliation{Max-Plank-Institute for Solid State Research, D-70569 Stuttgart, Germany}

\author{A. Sekiyama}
\affiliation{Graduate School of Engineering Science, Osaka University, Toyonaka, Osaka 560-8531, Japan}

\author{Y. Saitoh}
\affiliation{Materials Science Research Center, Japan Atomic Energy Agency, SPring-8, Hyogo 679-5148, Japan}


\begin{abstract}

In this study, we  systematically investigate 3D momentum($\hbar k$)-resolved electronic structures of Ruddlesden-Popper-type iridium oxides
Sr$_{n+1}$Ir$_n$O$_{3n+1}$ using soft-x-ray (SX) angle-resolved photoemission spectroscopy (ARPES).
Our results provide direct evidence of an insulator-to-metal transition that  occurs upon increasing the dimensionality of the IrO$_2$-plane structure.
This transition occurs when the spin-orbit-coupled  $j_{\rm eff}$=1/2 band changes its behavior 
in the dispersion relation and moves across the Fermi energy.
In addition, an emerging  band along  the $\Gamma$(0,0,0)-R($\pi$,$\pi$,$\pi$) direction is found to play a crucial role in the metallic characteristics of SrIrO$_3$.
By scanning the photon energy  over 350~eV,  we reveal the 3D Fermi surface in SrIrO$_3$ 
and   $k_z$-dependent oscillations of photoelectron intensity in Sr$_3$Ir$_2$O$_7$.
In  contrast to   previously reported results obtained using low-energy photons, 
folded bands derived from  lattice distortions and/or magnetic ordering make significantly weak (but  finite) contributions 
to the $k$-resolved photoemission spectrum.
At the first glance, this  leads to  the ambiguous result 
that the observed $k$-space topology is consistent with the unfolded  Brillouin zone (BZ) picture derived from a non-realistic simple  square or cubic Ir lattice.
Through careful analysis, we determine that a superposition of the folded and unfolded band structures has been  observed 
in the ARPES spectra obtained using  photons in  both ultraviolet and SX regions. 
To corroborate the physics deduced using low-energy ARPES studies, 
we  propose to utilize SX-ARPES as a powerful complementary technique, as this method  surveys more than one whole BZ and provides a panoramic view of electronic structures.

\end{abstract}

\pacs{74.25.Jb, 71.70.Ej, 71.20.-b, 71.30.+h}




\maketitle

%
%

\section{Introduction}

Spin-orbit coupling (SOC)  of an electron, a relativistic quantum effect, results in various exotic phenomena and has recently opened up new frontiers in solid state physics. 
Apart from the phenomena related to  surface science, such as the Rashba and quantum spin Hall effects~\cite{Bychokov84,Kane05},
there are some cases in which  SOC  plays a predominant role in the bulk nature of condensed matter.
Among them, Ruddlesden-Popper-type iridium oxides (RP-iridates) Sr$_{n+1}$Ir$_n$O$_{3n+1}$  offer an excellent opportunity for studying the interplay between the SOC and electron correlation effects in the bulk material, 
since the strength of the electron correlation changes together with the  dimensionality of the IrO$_2$-plane structure depending on the number  $n$~\cite{Moon08,Matsuno15}.

Two-dimensional layered  Sr$_2$IrO$_4$ ($n$=1; Sr214), which contains  an isolated  single IrO$_2$ plane, has the strongly spin-orbit-coupled $J_{\rm eff}$=1/2 state, in which
Ir $5d$ levels are  partially occupied by electrons (where $J_{\rm eff}$ stands for the {\it resultant} effective total angular momentum)~\cite{Kim08,Kim09}.
Following are some salient features of this state: 
(1) the SOC in the Ir 5$d$ electrons has an energy scale comparable to that of hopping integrals of 5$d$ electrons and the $d$-$d$ Coulomb interaction because of the large atomic number $Z$ (=77).
This leads to a significant splitting of Ir $5d$ $t_{2g}$  levels. 
The five 5$d$ electrons occupy the upper $j_{\rm eff}$=1/2 and lower $j_{\rm eff}$=3/2 states (where $j_{\rm eff}$ stands for  the {\it one-electron} effective total angular momentum). 
The $j_{\rm eff}$ states are represented as linear combinations of atomic $t_{2g}$ orbitals with a mixture of up and down 5$d$ electron spins;
(2) the half-filled $j_{\rm eff}$=1/2 band, which plays an important role in the unique nature of Sr214, further splits into two bands across the Fermi energy ($E_F$) because of antiferromagnetic and $d$-$d$ Coulomb interactions, whereas the $j_{\rm eff}$=3/2 bands are fully occupied and lie below $E_F$.
The presence of the half-filled insulating band (the so-called lower Hubbard band) indicates that Sr214 is  a promising candidate for the parent material of   high-$T_c$ superconductors, in analogy with La$_2$CuO$_4$~\cite{Watanabe13,Kim14,Torre15}.
In contrast, the final member of the system, SrIrO$_3$ ($n$=$\infty$; Sr113) shows   metallic behavior above $\sim$150~K~\cite{Kim06,Matsuno15}.
Sr113 is now attracting much attention because a bulk semi-metallic ground state with a new topological phase 
has been proposed on the basis of several  theoretical studies~\cite{Carter12,Zeb12,Okamoto13,Kim15,Chen15}.
In fact, the semi-metallic behavior of the Hall coefficient and possible Dirac-like linear dispersions in the electronic structures projected onto the 2D momentum (or wave number $k$) space were reported~\cite{Nie15,Matsuno15,Liu15}.
Between Sr214 and Sr113, an insulator-to-metal transition occurs with increasing  dimensionality.
The antiferromagnetic insulator Sr$_3$Ir$_2$O$_7$ ($n$=2; Sr327) is located close to the border of the transition~\cite{Carter13,Matsuno15}.

In the present study, we integrate the evolution of 3D electronic structures
with  the dimensionality of the IrO$_2$-plane structure, and the synergy effects that occur between the SOC and   electron correlations 
in  $J_{\rm eff}$ ground states  in the  RP-iridates Sr214, Sr327, and Sr113. 
To address these issues,  we use angle-resolved photoemission spectroscopy (ARPES) with  brilliant soft-x-ray (SX) synchrotron radiation.
In the SX region,   photoemission spectroscopy is suited to the investigation of electronic structures in the bulk owing to the large probing depth~\cite{Tanuma94,Sekiyama00,Yamasaki10}.
There are two more  benefits in using SX-ARPES for studying Ir $j_{\rm eff}$ states:
(1) a high Ir $5d$-O $2p$  sensitivity ratio, which is about 60 times higher than that in the vacuum ultraviolet (VUV) region~\cite{Lindau85} and
(2) a large survey area  that covers more than one whole Brillouin zone (BZ) in the 3D $k$ space.
This is made possible  by scanning a photoelectron acceptance angle of about 10$^\circ$ and  a photon energy of over 350~eV~\cite{Sekiyama04,Yano07}.

Consequently, we discover  $k$-resolved electronic structures in the RP-iridates consistent with a BZ derived from a simple  square or cubic Ir lattice,
in  contrast to the  previously reported UV and VUV-ARPES results~\cite{Kim08,Wang13,Moreschini14,Torre15,Nie15}.
Folded bands attributed to  lattice distortions and/or  magnetic ordering give significantly weak (but  finite) contributions 
to the  spectrum as a result of a matrix element effect dominant in the SX region.
Through careful analysis, we conclude  that a superposition of the folded and unfolded band structures has been  commonly observed 
in  ARPES experiments using   photons in  both UV and SX regions. 
Meanwhile, we  successfully observe the electronic structures unique to each iridate and find direct evidence for the insulator-to-metal transition.
The $j_{\rm eff}$=1/2 band  moves across  $E_F$ with  increasing iridate dimensionality.
Furthermore, an emerging  band along  the $\Gamma$(0,0,0)-R($\pi$,$\pi$,$\pi$) direction is shown to play a crucial role in the metallic behavior observed  in Sr113.

\begin{figure*}
\includegraphics[width=14.5cm,clip]{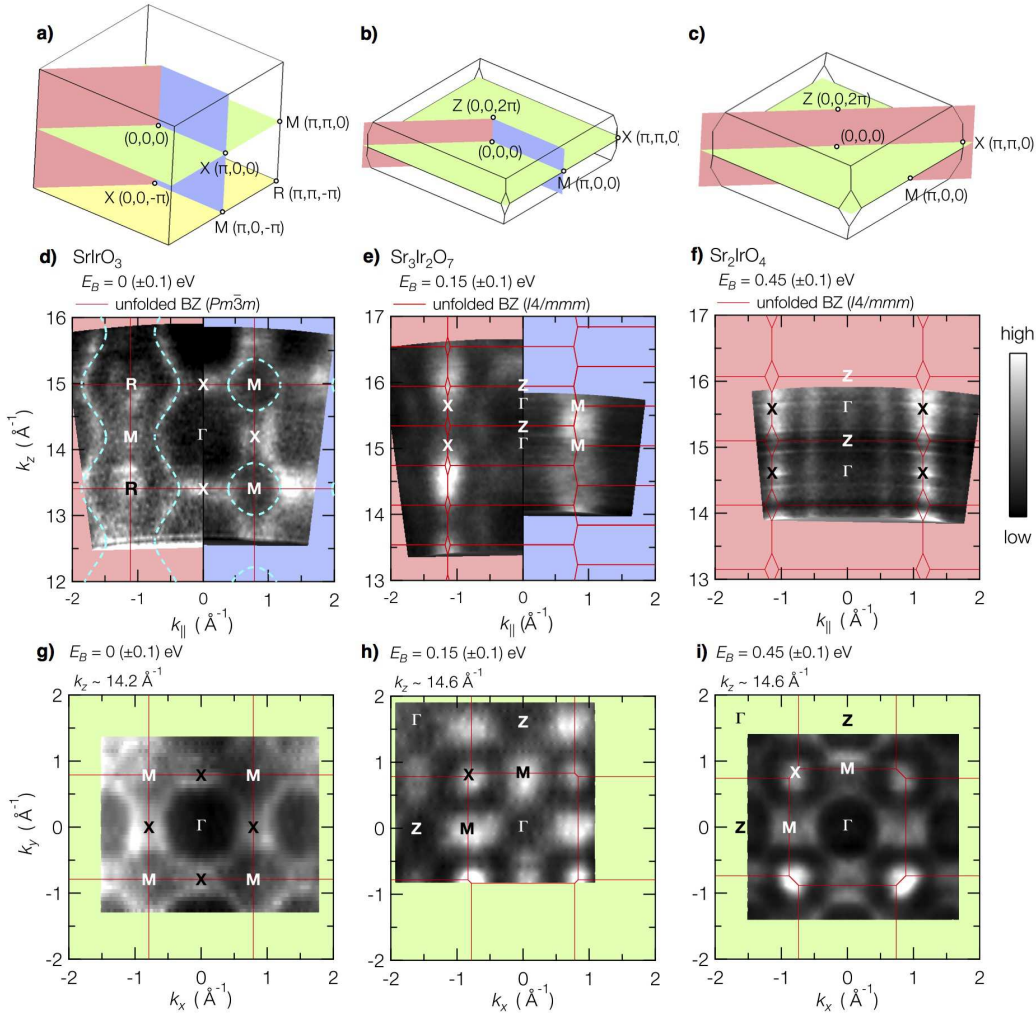}
\caption{
Constant-energy surfaces in the {$\boldsymbol k$} space obtained using SX-ARPES.
The first BZs of (a) Sr113, (b) Sr327, and (c) Sr214 for  RP-type crystal structures {\it without} consideration of the  IrO$_6$ octahedral rotation are shown.
Note that the lattice constants $a$ and $c$ are omitted from the coordinate points.
The constant-energy surfaces  in the $k_{||}$-$k_z$ and $k_x$-$k_y$ planes for Sr113 [(d) and (g)], Sr327 [(e) and (h)], and   Sr214 [(f) and (i)] are displayed.
Solid red lines indicate the cross-sectional view of the  BZs illustrated in (a)-(c).
Dashed curves in (d) show the cross-sectional FS discussed in Section III B.
The surfaces in (d)-(i) were obtained by integrating the momentum-distribution curves around a certain $E_B$  over an energy window of $\pm$100~meV~\cite{note0}.
Their surface topology is qualitatively similar to that with smaller energy windows under the present experimental conditions.
}
\label{Fig_1}
\end{figure*}

\section{experimental}

Thin films of pseudo-cubic perovskite-type  Sr113 were grown on  Nb(0.05~wt\%)-doped SrTiO$_3$ (001) substrates by  pulsed laser deposition (PLD).
The typical film thickness   and  out-of-plane (in-plane) lattice constant were determined to be 240~{\AA} and 3.985 (3.905)~{\AA}, respectively, 
using  x-ray diffraction (XRD) measurements,
indicating the in-plane epitaxial growth on the SrTiO$_3$ substrate   with the original cubic lattice constant of Sr113 in the out-of-plane direction.
Details of the sample preparation and  characterization methods have been reported elsewhere~\cite{Matsuno15}.

Single crystals of Sr214 and  Sr327 were  prepared using a flux method with SrCl$_2$ as the flux material.  
A mixture of the starting materials (SrCO$_3$ and IrO$_2$), together with the flux, was placed into Pt crucibles and heated to 
1350~$^\circ$C for Sr214 (1100~$^\circ$C for Sr327), and was maintained for 12 h.
Subsequently, the temperature was lowered to 900~$^\circ$C for 50 h. 
After cooling, plate-like crystals were obtained.
In both complexes,   the typical crystal size was about 0.8 mm$\times$0.6 mm$\times$0.3 mm. 
The crystal structures were evaluated using powder XRD at the beamline BL02B2 of SPring-8. 
In the case of Sr214, the lattice constants $a^\prime(=\sqrt{2}a)$ and $c^\prime(=2c)$, analyzed by Rietveld method
 based on the space group $I4_1$/{\it acd}, were 5.4955~{\AA} and 25.8130~{\AA}, respectively. 
Meanwhile, for Sr327, the lattice constants were determined to be  $a^\prime(=\sqrt{2}a)$=5.5157(2)~{\AA}, 
$b^\prime(\simeq\sqrt{2}a)$= 5.5188(2)~{\AA}, and $c$=20.9027(2)~{\AA} with the symmetry of {\it Bbeb}.
The temperature dependencies of the electrical resistivity and the magnetization under 1~T were measured  between 4 and 300~K
using a quantum design physical property measurement system (PPMS)  and  a magnetic property measurement system (MPMS). 
Sr214 (Sr327) samples displayed an insulating character across the entire temperature range and magnetically ordered behavior below 
the N\'eel temperature, $T_N$$\sim$240~K ($\sim$280~K), 
as reported previously~\cite{Crawford94,Cava94,Cao98,Cao02,Fujiyama12}.

SX-ARPES experiments were performed at the Japan Atomic Energy Agency (JAEA) actinide science beamline BL23SU of SPring-8 using the Gammadata-Scienta SES-2002 electron-energy analyzer and unpolarized light delivered by a twin helical undulator~\cite{Saitoh12}.
Films of Sr113  were prepared with and without a two-monolayer-thick SrTiO$_3$ capping layer.
These films were stored under a high-purity N$_2$ gas atmosphere  during travel from the PLD growth chamber to the load lock chamber in the beamline, and
then transferred to the ARPES chamber under  ultrahigh vacuum (UHV).
At the measuring temperature $T$=20~K, it was confirmed that both films  provided  qualitatively similar valence-band spectra in  the binding energy ($E_B$) range of 0-3~eV
in addition to quantitatively similar analytical results with respect to the topology of the Fermi surface (FS).
To obtain clean (001) surfaces for Sr214 and Sr327, single crystals were cleaved {\it in situ} in  UHV with a base pressure better than 1$\times$10$^{-8}$~Pa at the measurement  temperatures~\cite{Suppl}.
The energy resolution of the measurements of the energy-band dispersions along high-symmetry lines (for $k$-space maps) was set to  about 130 (250) meV 
at a typical  photon energy of h$\nu$$\simeq$750~eV.
The angular resolution was 0.2$^\circ$  (0.3$^\circ$) parallel  (perpendicular) to the analyzer slit, 
leading to an in-plane momentum resolution $\Delta k_{||}$ (or $\Delta k_x$,  $\Delta k_y$)  of, at most, 0.08~{\AA$^{-1}$}.
Meanwhile, the momentum resolution along the $k_z$ axis, $\Delta k_z$, was estimated to be 0.04~{\AA$^{-1}$} at the typical photon energy, which was better than one-third of  $\Delta k_z$ in the VUV region owing to the high bulk sensitivity in the SX region~\cite{Bulk_sensitivity}.
The emission angle of the photoelectron at each photon energy was converted into $k_x$ or $k_y$ ($k_z$) considering the photon momentum (and  inner potential $V_0$,
which was  experimentally estimated to be 10($\pm$3), 18($\pm$4), and 27($\pm$2)~eV for Sr113, Sr327, and Sr214, respectively)~\cite{Suppl}.
The Fermi energy was determined from the photoemission spectra of  {\it in situ} evaporated gold films.


\section{results and discussion}
\subsection{ Overall electronic structures in RP-iridates}

We  show the experimentally obtained 3D $k$-space maps of  the valence bands in   RP-iridates. 
First, note that the main features of  $k_x$-$k_y$ maps for all three iridates shown in Figs.~\ref{Fig_1}(g)-(i) are consistent with 
the unfolded BZ picture  [see also  Figs.~\ref{Fig_1}(a)-(c)] that were derived from crystal structures without  IrO$_6$ octahedral rotation and/or antiferromagnetic ordering.
As a result of  the rotation and magnetic ordering, the space group changes from $Pm\bar{3}m$ to $Pbnm$ for Sr113, from $I4/mmm$ to $Bbeb$ for Sr327, and from $I4/mmm$ to $I4_1/acd$ for Sr214, leading to band folding and  smaller BZs.
However, these folded bands were less visible or, in some cases, hardly seen in the SX-ARPES  (discussed later).
Meanwhile,  band folding in the $k_x$-$k_y$ plane has been clearly observed in previous UV and VUV-ARPES experiments 
for the iridates~\cite{Kim08,Wang13,Moreschini14,Torre15,Nie15}.
These observations, which probably result from a matrix element effect, indicate that the lattice distortion and magnetic phase transition only make slight or perturbative contributions to the photoemission spectrum  in the SX region.
Here, we show, in a sense,  the essential  electronic structures expected in ``undistorted'' RP-iridates,
whose crystal structures are similar to those of RP-ruthenates  and RP-cuprates.

Cross-sectional images of the FS  in Sr113 are shown in Figs.~\ref{Fig_1}(d) and (g).
The figures show that the $\Gamma$(0,0,0)-X($\pi$,0,0)-M($\pi$,0,$\pi$)-X(0,0,$\pi$) surface  in the $k_{||}$-$k_z$ plane is 
equivalent to the $\Gamma$(0,0,0)-X($\pi$,0,0)-M($\pi$,$\pi$,0)-X(0,$\pi$,0) surface in the $k_x$-$k_y$ plane, which is consistent with the cubic BZ.
This result  demonstrates that,  in the $k_z$ direction,    vital momentum broadening and  other matrix element effects were not observed.
Meanwhile,  in the case of Sr327, an incommensurate periodicity with the BZ along the $k_z$ axis is seen [Fig.~\ref{Fig_1}(e)], 
while a $k_z$-independent constant-energy surface has been found  for Sr214,
as can be expected from the  strong two-dimensionality of the electronic states derived from the single-layer perovskite structure.

\begin{figure}
\includegraphics[width=7.5cm,clip]{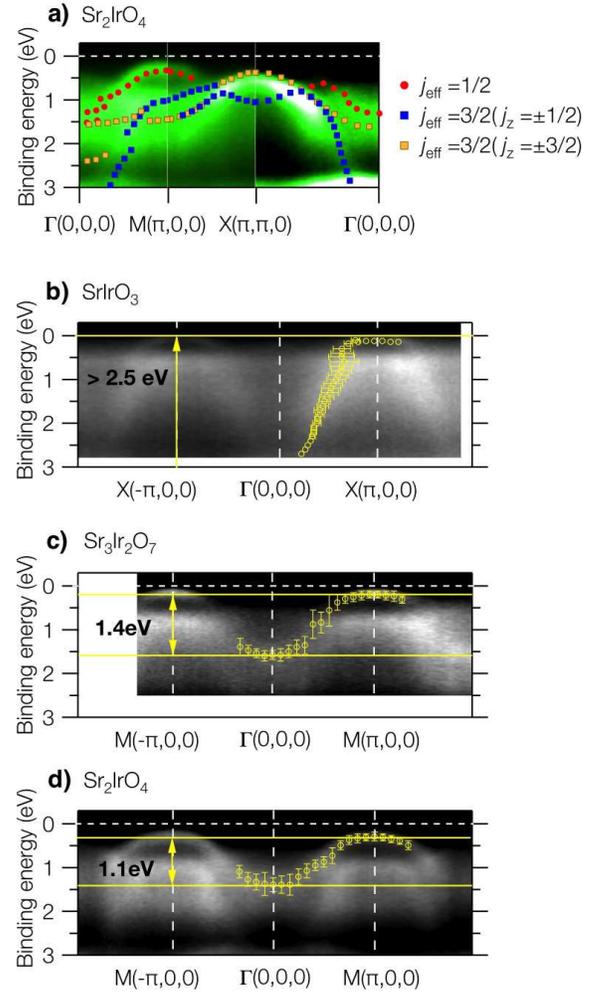}
\caption{
$j_{\rm eff}$=1/2-band width depending on the dimensionality of the IrO$_2$-plane structure.
(a) Energy bands seen in the SX-ARPES spectra of Sr214 and the schematic view of their  characters  assigned by calculations (see  Fig.\ref{Fig_6} for details).
The $j_{\rm eff}$=1/2-band dispersions along the (0,0,0)-($\pi$,0,0) line are shown for (b) Sr113, (c) Sr327, and (d) Sr214.
Open circles indicate the peak positions estimated from the momentum-(energy-)distribution curves of Sr113 (Sr327 and Sr214). 
The error bar represents statistical variability and the variation of the peak position between the original and the second-derivative images.
}
\label{Fig_2}
\end{figure}

\begin{figure*}
\includegraphics[width=11.5cm,clip]{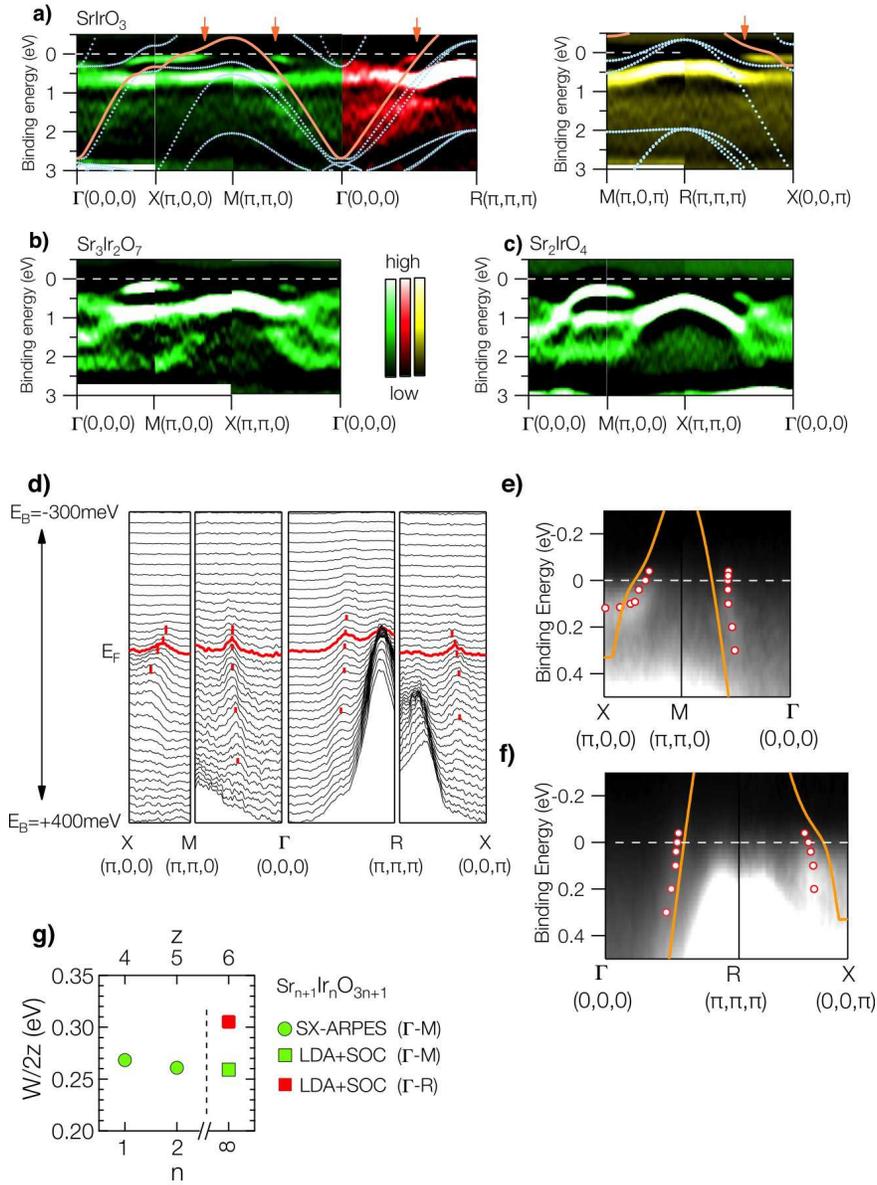}
\caption{
Energy-band dispersions along high-symmetry lines in (a) Sr113, (b) Sr327, and (c) Sr214.
To yield a better view of the dispersions, the second-derivative images are shown.
The color of the images corresponds to that of the BZ planes [Figs.~\ref{Fig_1}(a)-(c)], in which the high-symmetry line is included.
The arrows in (a) indicate the positions where  the observed bands go across  $E_F$ [see also (d)].
The LDA+SOC band structure of  cubic Sr113 is also shown in (a).
Solid orange lines indicate  the $j_{\rm eff}$=1/2 band.
(d) Momentum-distribution curves of Sr113 near $E_F$ along the high-symmetry lines, where the $j_{\rm eff}$=1/2 band goes across the $E_F$.
Peak positions for the momentum-distribution curves, estimated using a peak-fitting analysis, are also shown by  red bars~\cite{Suppl}.
(e) and (f) Enlarged band dispersions in the vicinity of $E_F$. 
Open circles in red indicate the peak positions in 
either the momentum- or energy-distribution curve.
Calculated $j_{\rm eff}$=1/2 bands are  shown by  solid  orange lines.
(g) $W$/2$z$ value corresponding to an averaged hopping integral $t$  in  the $j_{\rm eff}$=1/2 band, 
where $z$ stands for the number of  nearest-neighbor Ir atoms.
Here,  $W$ for Sr214 and Sr327 is  defined  as 2$w$, where $w$ is the $j_{\rm eff}$=1/2-band width in the occupied state.
The dashed line divides the insulating  and metallic states.
}
\label{Fig_3}
\end{figure*}

Turning to Fig.~\ref{Fig_2}, we show the evolution of  the  $j_{\rm eff}$-band width  with increasing  dimensionality of the IrO$_2$-plane structure.
The top and bottom of the $j_{\rm eff}$=1/2 band in Sr214 were located at M point ($\pi$,0,0) and $\Gamma$ point (0,0,0), respectively, as shown in Fig.~\ref{Fig_2}(a).
Upon changing  the structure of the IrO$_2$ planes from 2D (Sr214), through quasi-2D (Sr327), to 3D (Sr113), 
the $j_{\rm eff}$=1/2-band width increases, as  expected [Figs.~\ref{Fig_2}(b)-(d)].
To understand this in more detail,   the energy-band dispersions along the high-symmetry lines are shown for all  RP-iridates.
In the  in-plane  dispersions [green panels in Figs.~\ref{Fig_3}(a)-(c)],  the $j_{\rm eff}$=1/2  band in Sr113 goes across  the $E_F$
along  X($\pi$,0,0)-M($\pi$,$\pi$,0)-$\Gamma$(0,0,0) line for Sr113  as highlighted by arrows  [see also  Fig.~\ref{Fig_3}(d)],
and  forms a hole-like FS around the M point ($\pi$,$\pi$,0).
Meanwhile,  for Sr327 and Sr214,  the M($\pi$,0,0)-X($\pi$,$\pi$,0)-$\Gamma$(0,0,0) line corresponding to the same path in  Sr113
has its valence-band top at the M point ($\pi$,0,0) and  is folded at around ($\pi/2$,$\pi/2$,0), even though the photoelectron intensity emitted from the folded band is significantly weak [see also Fig.~\ref{Fig_2}(a) and \ref{Fig_6}(e)], creating a charge gap.
The  behavior of this band among the RP-iridates  demonstrates an insulator-to-metal transition that occurs with the increase in  dimensionality 
associated with moving from the quasi-2D (Sr327) to the 3D  (Sr113) compound.

By taking full advantage of the high-$k_z$ resolution available in  SX-ARPES, 
we  further investigated the energy-band dispersions for all other high-symmetry lines in the 3D BZ of Sr113.
Two additional $E_F$-crossing points were successfully observed in the $\Gamma$(0,0,0)-R($\pi$,$\pi$,$\pi$) and R($\pi$,$\pi$,$\pi$)-X(0,0,$\pi$) lines, 
as  marked by arrows  in the red and yellow panels in Fig.~\ref{Fig_3}(a) [see also  Fig.~\ref{Fig_3}(d)].
The observed dispersions were compared with the  results of the local density approximation (LDA)+SOC band structure calculations 
for  cubic Sr113 with $a$=3.985~{\AA}~\cite{Wien2k,note4}.
The behavior of the $j_{\rm eff}$=1/2 band near  the $E_F$-crossing points was  generally well-reproduced by the calculation shown in Figs.~\ref{Fig_3}(e) and (f);
however,  some inconsistency is apparent in Fig.~\ref{Fig_3}(a).
Some bands, whose tops are located at the R point ($\pi$,$\pi$,$\pi$), are pushed below $E_F$.
In contrast, other  bands around the $\Gamma$ point (0,0,0) seem to go above $E_F$.
These shifts make the  theoretically predicted hole (electron) pockets around the R ($\Gamma$) point disappear. 
Since our calculated results  do not have  any flat band in the M($\pi$,$\pi$,0)-$\Gamma$(0,0,0),  $\Gamma$(0,0,0)-R($\pi$,$\pi$,$\pi$), and $\Gamma$(0,0,0)-X($\pi$,0,0) lines, 
the observed bands at $E_B\simeq$ 0.7~eV around the $\Gamma$ point  are considered to be folded bands resulting from the IrO$_6$ octahedral rotation,  
as seen in the X($\pi$,$\pi$,0)-$\Gamma$(0,0,0) line for Sr327 and Sr214.
In fact, as mentioned above, the photoelectron intensity emitted from these bands was weaker than the original bands.
Other  bands that  have broad bandwidths of more than 2~eV are  hardly seen in Fig.~\ref{Fig_3}(a);
however, these bands become more visible following a change in  photon energy and  are  reproduced by our calculations~\cite{Suppl}.
Our results suggest that  the narrow $j_{\rm eff}$ bands observed in the low-energy ARPES experiments~\cite{Nie15,Liu15} are 
not derived from the electron correlation effect but from 
 lattice distortion together with the matrix element effect.
Unlike the results reported  in a previous study~\cite{Moon08},  we conclude that the electron correlation in Sr113  is rather weak.

\begin{table}
\caption{
The experimentally and theoretically obtained gradient of the $j_{\rm eff}$=1/2 band at $E_F$ and the Fermi velocity  in  Sr113.
}
\begin{ruledtabular}
\begin{center}
\begin{tabular}{lccc}

{} &high-symmetry line & $\displaystyle\frac{dE}{dk}\Big|_{k=k_F}$ & $v_F$   \\ 
{} & in cubic BZ & (eV\ {\AA}) & ($\times10^7$ cm/s) \\ \hline
\\
SX-ARPES & X -- M &  0.8 ($\pm$ 0.2)  &  1.2 ($\pm$ 0.3)  \\
 & M -- $\Gamma$ & $>$ 3.0   &  $>$ 4.6\\
 & $\Gamma$ -- R & 3.3 ($\pm$ 0.2)  &  5.0 ($\pm$ 0.3) \\
 & R -- X & 1.4 ($\pm$ 0.4)  &  2.2 ($\pm$ 0.6) \\
 \\
LDA+SOC  & X -- M &  0.64  &  0.97  \\
 & M -- $\Gamma$ & 2.5   &  3.8  \\
 & $\Gamma$ -- R & 3.1  & 4.7 \\
 & R -- X & 0.92  &  1.4 \\

\end{tabular}
\end{center}
\end{ruledtabular}
\label{Table_1}
\end{table}


\begin{figure}
\includegraphics[width=8.0cm,clip]{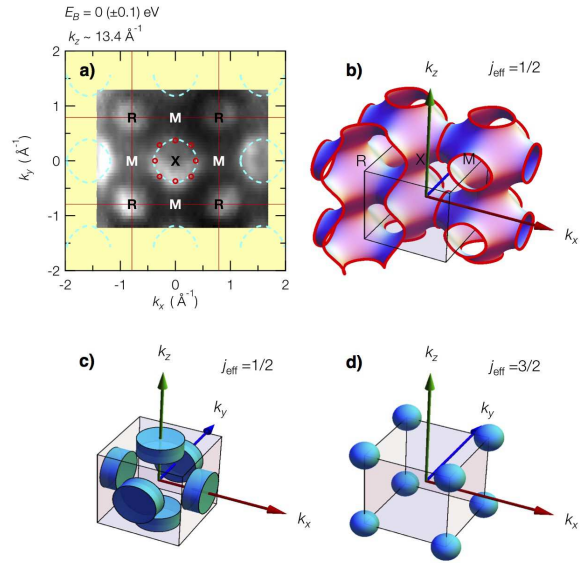}
\caption{
FS topology  of Sr113. 
(a) Constant-energy surface (including the FS) in the R($\pi$,$\pi$,$\pi$)-M($\pi$,0,$\pi$)-X(0,0,$\pi$)-M(0,$\pi$,$\pi$) plane.
Open circles in red indicate the  $k_F$ points estimated from the ARPES spectra~\cite{Suppl}. 
(b) Schematic image of a hole-like FS for a  simple cubic structure is shown. 
The cross-sectional views of the FS are also illustrated by the dashed line in (a) and Fig.~\ref{Fig_1}(d).
Electron reservoirs just below $E_F$ are  displayed in (c) and (d).
}
\label{Fig_4}
\end{figure}


\subsection{Fermiology of SrIrO$_{\bf 3}$}

%
Figures~\ref{Fig_3}(e) and (f) show the photoelectron-intensity distributions in the vicinity of $E_F$ around the M point ($\pi$,$\pi$,0) and R point ($\pi$,$\pi$,$\pi$).
It can be clearly seen that the $j_{\rm eff}$=1/2 band has various  slopes at the Fermi wavenumber $k_F$  on each high-symmetry line.
The estimated  Fermi velocities, $v_F$, are listed in Table~\ref{Table_1}.
The value near the ($\pi$/2,$\pi$/2,$\pi$) in  the R($\pi$,$\pi$,$\pi$)-X(0,0,$\pi$) line is almost equal to the previously 
reported value of  1.2~eV\ {\AA} at ($\pi$/2,$\pi$/2) in the surface-projected BZ~\cite{Nie15}.
Furthermore, we uncovered a significantly faster band  with 3.3~eV\ {\AA}, that is, $v_F$=5.0$\times10^{7}$cm/s 
near ($\pi$/2,$\pi$/2,$\pi$/2) in  the $\Gamma$(0,0,0)-R($\pi$,$\pi$,$\pi$) line; this value  is about half  the value generally reported for simple metallic systems~\cite{AM_pp38}.
These experimentally obtained $v_F$ values can be  reproduced by the LDA+SOC calculation for  cubic Sr113, also listed in Table~\ref{Table_1},
supporting the dominant $j_{\rm eff}$=1/2  character of this band in Sr113.

We now discuss the FS topology of Sr113 in  detail.
The  $j_{\rm eff}$=1/2-band  dispersions in Figs.~\ref{Fig_3}(e) and (f) suggest the presence of two hole pockets around the M point ($\pi$,$\pi$,0) and R point ($\pi$,$\pi$,$\pi$);
however,   Fig.~\ref{Fig_1}(d) shows that they are connected to each other along the R($\pi$,$\pi$,$\pi$)-M($\pi$,$\pi$,0) line, leading to a large open FS.
The experimental FS closely resembles a hole-like FS for a simple cubic structure in the tight-binding (TB) model  illustrated in Fig.~\ref{Fig_4}(b).
The FS is composed  of a simplified  energy band
$\epsilon_{\boldsymbol k}=-2t(\cos k_xa+\cos k_ya+\cos k_za)$,
where $t$ is an  hopping integral~\cite{CK_pp244}.
At $\epsilon_{\boldsymbol k}=0$,  the energy band reproduces a characteristic  FS topology 
in the R($\pi$,$\pi$,$\pi$)-M($\pi$,$\pi$,0)-$\Gamma$(0,0,0)-X(0,0,$\pi$) plane as shown in Fig.~\ref{Fig_1}(d) and 
the TB-model FS is consistent with the number of  $E_F$-crossing points observed in all high-symmetry lines.
Here, we roughly estimate the  number of the hole $n_h$,  not  using the LDA+SOC  but  the TB band
since the experimental  $k_F$ values are well reproduced by the TB-model calculation rather than the  LDA+SOC calculation~\cite{Suppl}.
Approximating the experimental FS by  the TB-model FS, $n_h$
in the Wigner-Seitz cell can be analytically obtained by
\begin{equation}
n_h=2\frac{V_{\rm FS}}{\displaystyle\bigg(\frac{2\pi}{a}\bigg)^3}=1,
\end{equation}
where $V_{\rm FS}$ (=4$\pi^3$/$a^3$) is the volume of the TB-model FS.
Following are the two principal results of the SX-ARPES: (1) the observed band  going across $E_F$ corresponds to $j_{\rm eff}$=1/2  and (2) $n_h$$\simeq$1.
These results suggest that 
Sr113 has a well-defined $J_{\rm eff}$=1/2 ground state, as do Sr214 and Sr327.
Thus, the $J_{\rm eff}$=1/2 picture is valid independent of the dimensionality  in RP-iridates. 
Here,  note that the FS consisting of only  hole-like bands appeared to be inconsistent with the strongly temperature-dependent 
and negative Hall coefficient and other theoretical predictions which suggest
that Sr113 is semi-metallic~\cite{Matsuno15,Carter12,Zeb12}.
In fact, it has been reported that  the electron-like FS is composed of some folded bands invisible in SX-ARPES~\cite{Nie15}.
Apart from that, we  observed a finite photoelectron intensity at $E_F$ around the R and X points,
as shown in Fig.~\ref{Fig_4}(a) [see also Figs.~\ref{Fig_1}(d) and (g)], resulting from the $j_{\rm eff}$=3/2 and 1/2 bands approaching  $E_F$ and a finite energy resolution.
These bands form  ``electron reservoirs" as shown in Figs.~\ref{Fig_4}(c) and (d), 
 and can contribute to the electron conductivity at high temperature.

\subsection{Bilayer coupling effect in Sr$_{\bf 3}$Ir$_{\bf 2}$O$_{\bf 7}$}

\begin{figure}
\includegraphics[width=8.6cm,clip]{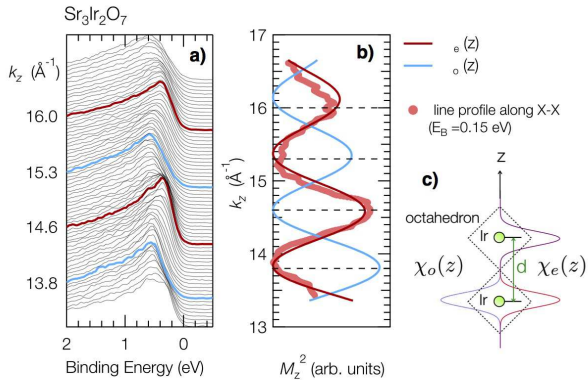}
\caption{
Bilayer splitting and  matrix element effect in Sr327.
(a) Energy-distribution curves along the X-X line in the $k_z$ direction.
(b) Transition probability along the $k_z$ axis, $M_z^2$, induced by photoexcitation for a bilayer-coupled system.
The photoelectron intensity  along the X-X line from Fig.~\ref{Fig_1}(e) is also shown.
(c) Schematic illustration of the bilayer-coupled wave functions.
}
\label{Fig_5}
\end{figure}

Let us now discuss  the  dependence of  the photoelectron intensity on $k_z$ in the constant-energy surface of Sr327.
As has been  shown in Fig.~\ref{Fig_1}(e), the electronic structures along the $k_z$ axis extend beyond the BZ.
This result demonstrates  that the electronic states predominately contributing to  photoemission originate from an internal structure of the unit cell, 
that is, the  IrO$_2$ bilayer  unique to Sr327.
To understand the incommensurate periodicity of the electronic structures, we  examined the matrix element effect derived from the bilayer coupling 
in a manner analogous to that described for 
Bi$_2$Sr$_2$CaCu$_2$O$_{8+\delta}$ (Bi2212)~\cite{Feng02}.
We assumed that the transition matrix element $M$ for  one-particle excitation can be  separated into two parts, {\it i.e.}, $M = M_{||}+M_{z}$,
where $M_{||}$ and $M_z$ are the  in-plane and out-of-plane components of $M$, respectively.
Furthermore, given that  the initial-state wave function in an Ir ion can be simplified to be a single Gaussian function~\cite{Feng02},
the bilayer-coupled states are expressed by a linear combination of these Gaussian functions, that is,  
the odd and even functions $\chi_o$ and $\chi_e$, as shown in Fig.~\ref{Fig_5}(c).
The matrix element $M_z$  in the electric dipole transition to a free-electron final state is now given  by
\begin{widetext}
\begin{eqnarray}
M_{z}(k_z)&=&\int_{-\infty}^{\infty}z\exp\Bigg[-\frac{\displaystyle\bigg(z-\frac{d}{2}\bigg)^2}{(\beta d)^2}\Bigg]\exp(ik_zz)dz\pm\int_{-\infty}^{\infty}z\exp\Bigg[-\frac{\displaystyle\bigg(z+\frac{d}{2}\bigg)^2}{(\beta d)^2}\Bigg]\exp(ik_zz)dz\\
 \nonumber \\ 
&=&
\begin{cases}
i\sqrt{\pi}\beta d\exp\bigg(\displaystyle-\frac{\beta^2d^2}{4}k_z^2\bigg)\Bigg\{d\sin \bigg(\frac{d}{2}k_z\bigg)+\beta^2d^2k_z\cos\bigg(\displaystyle\frac{d}{2}k_z\bigg)\Bigg\}\ \ :{\rm for\ }\ \chi_{e}(z),\\ \nonumber
\\  \nonumber
\sqrt{\pi}\beta d\exp\bigg(\displaystyle-\frac{\beta^2d^2}{4}k_z^2\bigg)\Bigg\{d\cos \bigg(\frac{d}{2}k_z\bigg)-\beta^2d^2k_z\sin\bigg(\displaystyle\frac{d}{2}k_z\bigg)\Bigg\}\ \ \ :{\rm for\  }\ \chi_{o}(z),\\   
\end{cases}\\  \nonumber
\end{eqnarray}
\end{widetext}
where $d$ is the distance between  IrO$_2$ layers, 4.070{\AA}~\cite{Cao02}.
The damping parameter $\beta$ was set to be 1/35 so as to reproduce the experimental results.
Note that this optimal value of $\beta$ is much smaller than that reported for Bi2212 ($\beta$=1/6) in the VUV-ARPES experiment,
indicating  the narrow spread of the wave function in the present initial state~\cite{Feng02}.
In the photoemission event  using  x-rays,  the valence electron is considered to be emitted spatially close to the ionic core, which can result in small $\beta$ values~\cite{Suga09}.

As shown in Fig.~\ref{Fig_5}(b), the oscillation of the photoelectron intensity at $E_B$=0.15~eV 
along the X-X line in the  $k_z$ direction  is well-reproduced by the calculated $M_z(k_z)$.
This demonstrates  that the oscillation near $E_F$ is derived from  the  symmetrically coupled Ir $5d$ state~\cite{note1}.
The energy distribution curves along the X-X line are shown in Fig.~\ref{Fig_5}(a).
It was found that the peak contributing to the oscillation shifts to a point corresponding to a deeper $E_B$  when the transition probability $M_z^2$ 
for the anti-symmetrically coupled wave function $\chi_o(z)$ becomes large, as shown in Fig.~\ref{Fig_5}(b).
Therefore, the two peaks at $E_B\simeq$350 and 550~meV can be assigned to the anti-bonding and bonding states, originating from   $\chi_e(z)$ and  $\chi_o(z)$, respectively.
This  is known as  bilayer splitting, and the splitting energy is estimated to be about 200~meV.
Bilayer splitting in the folded band was observed at the $\Gamma$ point~\cite{Moreschini14}.
Interestingly,  with changing $k_z$ values, the strengths of the photoelectron intensities along  X-X and M-M lines increase and decrease alternately [see  Fig.~\ref{Fig_1}(e)], 
indicating that the anti-bonding state at the M point is derived from  $\chi_o$ instead of $\chi_e$.
Considering, as in the case of Sr214, that the $j_{\rm eff}$=1/2 (3/2) band  predominates  in the vicinity of $E_F$ at the M (X) point, it is clear  that the nature of bilayer coupling  depends on the $j_{\rm eff}$ band character.

\subsection{Superposition of folded and unfolded  band structures in Sr$_{\bf 2}$IrO$_{\bf 4}$}

\begin{table}
\caption{
The number of $j_{\rm eff}$ bands at each high-symmetry point between $E_F$ and $E_B$=2eV    in Sr214,
observed using  ARPES experiments at various photon energies.
}
\begin{ruledtabular}
\begin{center}
\begin{tabular}{llcccc}
 & & (0,0) & ($\pi$,0) & ($\pi$/2,$\pi$/2) & ($\pi$,$\pi$)   \\
h$\nu$  &  $I4$/{\it mmm}  &$\Gamma$ & M &  &  X  \\  
 & $I4_1$/{\it acd}&$\Gamma$ & X & M &   \\ \hline
  760 eV & & 3 & 3 & 2 & 1\\
 85 eV\footnote{Ref. [\onlinecite{Kim08}].} & & 1 & 2 & 2 & - 
  \\
 21.2 eV\footnote{Ref. [\onlinecite{Uchida14}].} & &  3 & 3 & 2 & 2  
\end{tabular}
\end{center}
\end{ruledtabular}
\label{Table_2}
\end{table}


\begin{figure}
\includegraphics[width=7.5cm,clip]{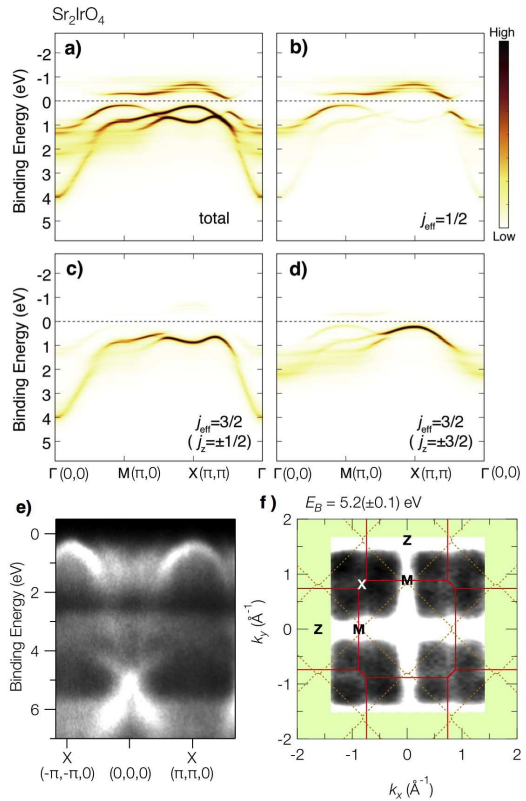}
\caption{
(a) Theoretical energy spectrum based on the three-orbital Hubbard model with consideration for the unfolded BZ of Sr214.
(b)-(d) Spectra projected onto the atomic $j_{\rm eff}$=1/2, 3/2 ($j_z$=$\pm$1/2), and 3/2 ($j_z$=$\pm$3/2) bases, respectively. 
The parameters and their values for the calculation are identical to those used in the previous study
({\it i.e.}, $U$=1.44~eV, $U^{\prime}$=1.008~eV, $J$=0.216~eV, and $\lambda$=0.432~eV in Ref.~[\onlinecite{Watanabe10}]).
(e) Experimental energy-band dispersion along  $\Gamma$(0,0,0)-X($\pi$,$\pi$,0) line. 
(f) Constant-energy surface at $E_B$=5.2~eV.
These images were taken at h$\nu$=760~eV.
In (f),   unfolded and folded BZs are  drown as solid and dotted lines, respectively.
}
\label{Fig_6}
\end{figure}

Here, we  provide deeper insight into the inconsistency in the expected BZ size determined by  SX-ARPES and the other two experiments, UV-ARPES and XRD,
by discussing  the well-known electronic structure in Sr214.
As has already been reported, the crystal structure of Sr214 is determined by  XRD~\cite{Crawford94}.
The in-plane rotation (11$^\circ$) of the IrO$_6$ octahedra makes the primitive vectors rotate  and the in-plane (out-of-plane) lattice
constant $a$ ($c$) expands by a factor of $\sqrt{2}$ (2), changing the crystallographic space group from $I$4/{\it mmm} to $I4_1$/{\it acd}.
As a result, Sr214 has a small folded BZ unlike that of the triplet superconductor Sr$_2$RuO$_4$.
In comparison with the energy-band dispersions  at h$\nu$=760~eV  in  Fig.~\ref{Fig_3}(c) and at 21.2~eV in Ref.~[\onlinecite{Uchida14}],
the overall characteristic feature of the $j_{\rm eff}$ bands, including the number of  bands passing through the high-symmety points listed in Table~\ref{Table_2},
are in good agreement with  each other.
We stress that  the experimental results at both photon energies can be well-reproduced by the energy spectrum calculated  for the {\it unfolded} BZ  shown in Figs.~\ref{Fig_6}(a)-(d),
with the expectation of the band approaching  $E_F$ near the $\Gamma$ point.
Meanwhile, the other band, whose bottom  is located at  ($\pi$,$\pi$) near $E_B$=1~eV  at h$\nu$=21.2~eV, 
is not resolved at 760~eV but theoretically predicted within the unfolded BZ picture, as shown in Fig.~\ref{Fig_6}(c)~\cite{note2}.
It can be seen that the exceptional band near the $\Gamma$ point is derived from the band folding since it appears in the calculated spectrum for the folded BZ~\cite{Watanabe10}.
Nevertheless, the major feature in the $k$-resolved photoemission spectra obtained using both high- and low-energy photons are 
correctly explained  within the unfolded BZ picture in Sr214 as in other RP-iridates.
This  shows that, in any ARPES result,   the high-symmetry point  ($\pi$,$\pi$) should be distinguished from the $\Gamma$ point (0,0), 
even though they are identical in the surface-folded BZ~\cite{note3}.

Furthermore, it is clear that the photoelectron-intensity ratio of the folded band relative to the other unfolded bands 
does not change over the range of h$\nu$$\simeq$400 and 800~eV~\cite{Suppl}.
This indicates that either  the folded or unfolded band  is not derived from the surface structures
since the probing depth changes from about 9 to 15 {\AA} with increasing  photon energy.
We observed another folded band around the X point ($\pi$,$\pi$), whose intensity   is significantly weak, 
in the O $2p$ (and Ir $5d$ bonding) states, as shown in Fig.~\ref{Fig_6}(f).
Note that  the unfolded band structure in the O $2p$ states  remains  dominant,  even though  the oxygen atoms  change their positions due to the IrO$_6$ octahedral rotation.
Judging from the above results, we can conclude that the weak intensity arising from the folded bands in the SX-ARPES spectra is not caused  
by  the bulk/surface  or any specific band character.
We propose here that the most promising candidate for the origin of this folded/unfolded problem is  
the matrix element effect of the electric dipole transition in the photoemission process excited by  high-energy photons.
A better understanding of this phenomenon is expected to result from further theoretical studies.

\subsection{Structural dimensionality and metal-insulator transition}

\begin{figure}
\includegraphics[width=9.0cm,clip]{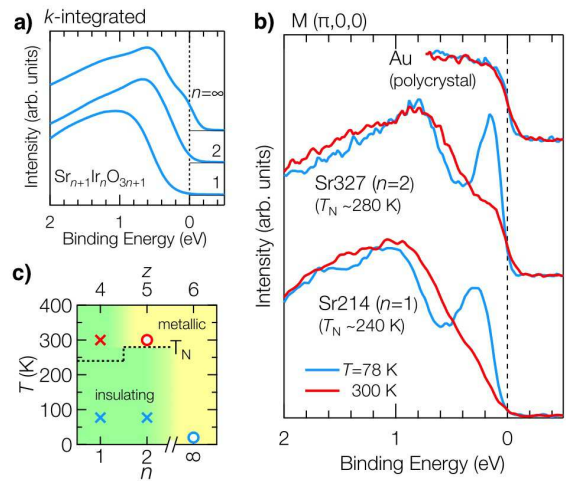}
\caption{
Metal-insulator transition in RP-iridates.
(a) $k$-integrated spectra of Sr113 ($n$=$\infty$), Sr327 ($n$=2), and Sr214  ($n$=1) obtained by  summing up 
the SX-ARPES spectra in both  $k_x$-$k_y$  and $k_{||}$-$k_z$ planes  shown in Figs.~\ref{Fig_1}(d)-(i).
(b) Temperature dependence of the SX-ARPES spectrum at the M point ($\pi$,0,0).
(c) The structural dependence of the electronic nature and its temperature variation,
where $z$ stands for the number of  nearest-neighbor Ir atoms.
Open circles (cross marks) indicate the temperatures at which the Fermi cutoff was (not) observed.
}
\label{Fig_7}
\end{figure}

Finally, we discuss the relationship among the electronic nature, structural dimensionality, and magnetic ordering in RP-iridates.
Based on a simple TB model, the hopping integral $t$ is expressed as $W$/2$z$, 
where $W$ and $z$ are the bandwidth and the number of  nearest-neighbor Ir atoms, respectively.
Figure~\ref{Fig_3}(g) shows a plot of the  $W$/2$z$ values for the $j_{\rm eff}$=1/2 band in the $\Gamma$(0,0,0)-M($\pi$,$\pi$,0) and $\Gamma$(0,0,0)-R($\pi$,$\pi$,$\pi$) lines 
obtained from the LDA+SOC calculation for Sr113.
The values in the $\Gamma$(0,0,0)-M($\pi$,0,0) line of Sr214 and Sr327, estimated from the dispersions observed  in Figs.~\ref{Fig_2}(c) and (d), are also shown.
The hopping integral in Sr113 along the $\Gamma$(0,0,0)-M($\pi$,$\pi$,0) line is nearly the same as the in-plane integrals in Sr214 and Sr327.
Meanwhile, the integral has a significantly large value along the $\Gamma$(0,0,0)-R($\pi$,$\pi$,$\pi$) line, 
in which the hopping is only possible in the 3D crystal structure unique to  Sr113.
This fact strongly suggests that the hopping of the carrier  along the body diagonal  direction plays a crucial  role in the metallic nature of Sr113.

In  contrast to Sr113 ($n$=$\infty$), Sr327 ($n$=2) and Sr214 ($n$=1) have no clear Fermi cutoff in the $k$-integrated spectra shown in Fig.~\ref{Fig_7}(a).
This feature becomes more apparent at the M point ($\pi$,0,0) where the $j_{\rm eff}$=1/2 band has the valence-band top in both compounds.
Figure~\ref{Fig_7}(b) shows that   the $k$-resolved spectrum of Sr327 has a prominent peak at  $E_B$$\simeq$150~meV below the magnetic ordering  temperature $T_N$,
whose tail goes across $E_F$ to  unoccupied states because of a finite energy resolution.
Even though non-zero photoelectron intensity  at $E_F$ was observed, 
the clear Fermi cutoff could not be found in the spectrum.
The peak observed below $T_N$  becomes wider and shifts to a deeper $E_B$ with decreasing $n$, {\it i.e.}, the structural dimensionality.
The peak shift results  in the creation of the charge gap in Sr214.

The $k$-resolved spectra of  Sr327 and Sr214 above $T_N$, also shown in Fig.~\ref{Fig_7}(b), resemble the $k$-integrated spectra of these compounds  in Fig.~\ref{Fig_7}(a), respectively, 
suggesting that  a $k$-broadening effect predominates at room temperature~\cite{note5}.
The $k$ broadening observed here  is derived from the matrix element effect mainly related to the Debye-Waller factor~\cite{Braun13}.
In addition, the energy broadening of a few hundred~meV resulting from the phonon-induced effect should be considered~\cite{Shevchik77}.
Despite these broadening effects, it is  revealed that the $k$-resolved spectra of Sr327 and Sr214 in the vicinity of $E_F$ are significantly different from each other.
The Fermi cutoff  in the spectrum of Sr327 appears above $T_N$,  indicating that an insulator-to-metal transition occurs across $T_N$.
Collapse of the charge gap  consistent with the drop of the electrical resistivity above $T_N$ demonstrates that a strong Mott insulating picture is not applicable to Sr327~\cite{Fujiyama12,King13,Carter13_2}.
In contrast, Sr214 has no Fermi cutoff in the $k$-resolved spectrum and  maintains the insulating character even above $T_N$.
Based on these results, we construct  a phase diagram in which 
the metallic nature of RP-iridates, mediated by the electrons in the spin-orbit coupled $j_{\rm eff}$=1/2 band,  develops step-by-step with increasing  dimensionality of the IrO$_2$-plane structure
and with the help of  magnetic ordering  as summarized in Fig.~\ref{Fig_7}(c).

\section{summary}

We measured the  3D  $k$-resolved electronic structures of three  RP-iridates Sr113, Sr214, and Sr327, 
using bulk-sensitive SX photoemission spectroscopy in conjunction with  theoretical calculations.
The most striking results in the present study are  as follows: 
(1) it was clearly demonstrated that the insulator-to-metal transition is induced by the variation in the $j_{\rm eff}$=1/2 band dispersion, which occurs  
with the increasing  dimensionality in the RP-iridates.
In addition, Sr327 solely showed the transition  in conjunction with the magnetic phase transition induced by temperature change.
(2) We   discovered a perfect 3D FS.
This was made possible by the use of the synchrotron light in the scanned photon energy range of more than 350~eV.
Furthermore, the dispersions among all the high-symmetry points in the cubic BZ  were experimentally revealed.
(3) Broadly dispersive $j_{\rm eff}$=1/2 bands, which only go across $E_F$ and are well-separated from  $j_{\rm eff}$=3/2 bands,
can be used to categorize  Sr113 as a $J_{\rm eff}$=1/2 itinerant electron system.
(4) We  directly observed  the photoelectron-intensity oscillation derived from the bilayer coupling  and found  its anisotropy in the {$\boldsymbol k$} space  
depending on the $j_{\rm eff}$ band character.

Note that the superposition of  the folded and unfolded band structures can be intrinsically observed 
in the ARPES experiments using any photons from the UV to the SX region.

\begin{acknowledgments} 

We would like to thank Y. Higa for supporting the SX-ARPES experiments, A.~Higashiya for supporting analysis, S. Suga for careful reading of the manuscript, H.~Wadati for fruitful discussions, and C.~Moriyoshi and Y.~Kuroiwa for the XRD measurements of Sr214 and Sr327. This work was performed under the Shared Use Program of JAEA Facilities (Proposals No. 2014A-E29, 2015A-E23, and 2015B-E23) with the approval of Nanotechnology Platform project supported by the Ministry of Education, Culture, Sports, Science and Technology (MEXT) (Proposals No.~A-14-AE-0022, No.~A-15-AE-0021) and JSPS Grant-in-Aid for Scientific Research (C)(No.~JP15K05186). 
H.W., T.S., and S.Y. was supported by Grant-in-Aid for Scientic Research from MEXT Japan under the Grant No. 25287096).
The synchrotron radiation experiments were performed at the JAEA beamline BL23SU (Proposals No.~2014A3882, No.~2015A3882, and No.~2015B3882) and the JASRI beamline BL02B2 (No.~2016A1230) of SPring-8.

\end{acknowledgments}


\appendix


\section{Detailed experimental conditions}
The detailed experimental conditions used in the soft-x-ray (SX) angle-resolved photoemission spectroscopy (ARPES) 
 of Ruddlesden-Popper-type iridium oxides (RP-iridates) are summarized in Table~\ref{Table_S1}.
To obtain the electronic structures in the $k_x$-$k_y$ plane,  emission-angle($\xi$)-dependent  spectra were measured
 at  a fixed angle $\theta$,
where  $\xi$ is defined as the angle formed between the direction directly toward the electron-energy  analyzer 
and  the emission direction of a  photoelectron detected through the entrance slit of the analyzer,
and $\theta$ is the angle between the directions toward the analyzer and normal to the sample surface.
The analyzer was designed such that the range of $\xi$ was set to be $\pm$6$^\circ$.
The dataset of the photoelectron intensity $I(\xi,\theta,E_k)$ was acquired by scanning  $\theta$, and then, converted to, for instance,  $I(k_x,k_y,E_B)$
using the following formulae:
\begin{eqnarray}
k_x&=&\frac{\sqrt{2mE_k}}{\hbar}\sin\xi,\\
k_y&=&\frac{\sqrt{2mE_k}}{\hbar}\sin\theta-q_{||},\\
E_B&=&h\nu-W-E_k,
\end{eqnarray}
where $E_k$, $E_B$, $h\nu$,  and $W$ indicate   the kinetic energy of the photoelectron, the binding energy of the emitted electron in the initial state, 
the photon energy,  and the work function, respectively.
 $q_{||}(=|{\boldsymbol q}|\cos{45^\circ}$ in the present study) is the photon momentum parallel to the sample surface.
 Here, we assume the free-electron final state.
Meanwhile,  the electronic structures along the $k_z$ axis were obtained by scanning the excitation photon energy. 
The dataset corresponding to $I(\xi,h\nu,E_k)$ at a fixed angle $\theta$ was converted to, 
for instance, $I(k_x,k_z,E_B)$ by Eqs.(A1)-(A3), and 
\begin{eqnarray}
k_z&=&\sqrt{\frac{2m(E_k+V_0)}{\hbar^2}-(k_x^2+k_y^2)} +q_\perp,
\end{eqnarray}
where  $q_{\perp}$ is the photon momentum perpendicular to the sample surface, and $V_0$ is the inner potential, which is a fitting parameter used in the  present study.

\begingroup      
%
\begin{table*}
\caption{
Detailed  conditions of SX-ARPES experiments for RP-iridates.
Regarding the sample configuration, the axis of the $\theta$ rotation is parallel to the entrance slit of the electron-energy analyzer.
$\theta$=0 indicates the normal emission of the photoelectrons.
Note that  it does not correspond to  $k_x$ (and/or $k_y$) =0 because of the momentum transfer from a photon to the photoelectron.
$\phi$ and $\psi$ are  the azimuth and tilting angles of the sample stage.
}
\begin{ruledtabular}
\begin{center}
\begin{tabular}{lccccc}
Compound & Mapping or & Photon energy   & $\theta$, $\phi$, $\psi$ & Temperature & Displayed in\\ 
 &   band dispersion &  (eV)  & ($^\circ$) & (K)  &  Figure(s)\\ \hline
SrIrO$_3$  & $\Gamma$-X-M-X  & 730 &  --5.0 to +6.0, 0, 0 & 20 & 1(g) \\
(without capping layer)  &  $\Gamma$-X & 730  & --0.7, 0, 0 & 20 & 2(b), 3(a) \\
   &  X-M & 730  & --3.8, 0, 0 & 20 &  11(b) \\
& X-M-R-M   & 650 &   --5.0 to +6.0, 0, 0  & 20 & 4(a) \\
&  X-M & 650 &   --1.0, 0, 0  & 20 & 3(a), 3(d), 3(e), 10(b),  11(a)\\
&  M-R & 650 &   --4.1, 0, 0  & 20 & 3(a) \\
 &   $\Gamma$-X-M-X &  575-910  &   --0.5, 0, 0  & 20 & 1(d), 8(a) \\
 \\
SrIrO$_3$  & $\Gamma$-M & 730 &   2.0, 45, 0  & 20 & 3(a), 3(d), 3(e),  10(a)\\
(with capping layer)  & $\Gamma$-R & 880 &  --5.3, 35, 45 & 20 & 3(a), 3(d), 3(f), 10(d), 10(e),  10(f) \\
 & $\Gamma$-R & 880 &  --7.0 to --1.0, 35, 45 & 20 & 10(g) \\
  & X-R & 810&  2.0, 45, 0 & 20 & 3(a), 3(d), 3(f), 10(c) \\
  &  $\Gamma$-X-R-M & 550-900 & 2.0, 45, 0  & 20 &1(d), 10(e) \\
  \\
  Sr$_3$Ir$_2$O$_7$  & $\Gamma$-M-X-M & 769 & --6.0 to +5.0, 0, 0  & 100 & 1(h) \\
  & $\Gamma$-M & 845 & --1.0, 0, 0  & 100 & 2(c), 3(b) \\
    & M-X & 845 & --3.7, 0, 0  & 100 & 3(b) \\
  & $\Gamma$-X & 855 & 0.6\footnote{These data were acquired from different crystal grains  in a series of measurements.}, 45, 0  & 100 & 3(b) \\  
    & $\Gamma$-X-X-Z & 650-995 &  --0.6$^a$, 45, 0 & 100 & 1(e), 5(a), 5(b),  8(b) \\  
  & $\Gamma$-M-M-Z & 710-900 &  --1.0, 0, 0 & 100 & 1(e) \\  
    & M& 689 &  --1.0, 0, 0 & 78, 300 & 7(b) \\  
    \\
  Sr$_2$IrO$_4$  & $\Gamma$-M-X-M & 760 &  --6.3 to 5.1, 0, 0 & 78 & 1(i), 6(f) \\
   & $\Gamma$-M  & 760 & --0.7, 0, 0  & 78 & 2(a), 2(d), 3(c) \\
      &  M-X & 760 & --3.9, 0, 0  & 78 &  2(a), 3(c) \\
      & $\Gamma$-X & 405 &  --1.0, 45, 0 & 100 & 2(a), 3(c), 12(a),  12(e) \\
      & $\Gamma$-X & 760 & --0.8, 45, 0  & 100 & 6(e), 12(b),  12(f) \\
            & $\Gamma$-X & 812 & --1.0, 45, 0  & 100 & 12(c), 12(g) \\
                  & $\Gamma$-X & 838 & --0.8, 45, 0  & 100 & 12(d), 12(h) \\
   & $\Gamma$-X-X-Z & 690-900 & --1.0, 45, 0  & 100 & 1(f), 8(c) \\
       & M& 760 &  --0.7, 0, 0 & 78, 300 & 7(b) \\  
\end{tabular}
\end{center}
\end{ruledtabular}
\label{Table_S1}
\end{table*}
\endgroup 

\section{Estimation of inner potential {\emph V}$_{\bf 0}$}

\begin{figure*}
\includegraphics[width=14.5cm,clip]{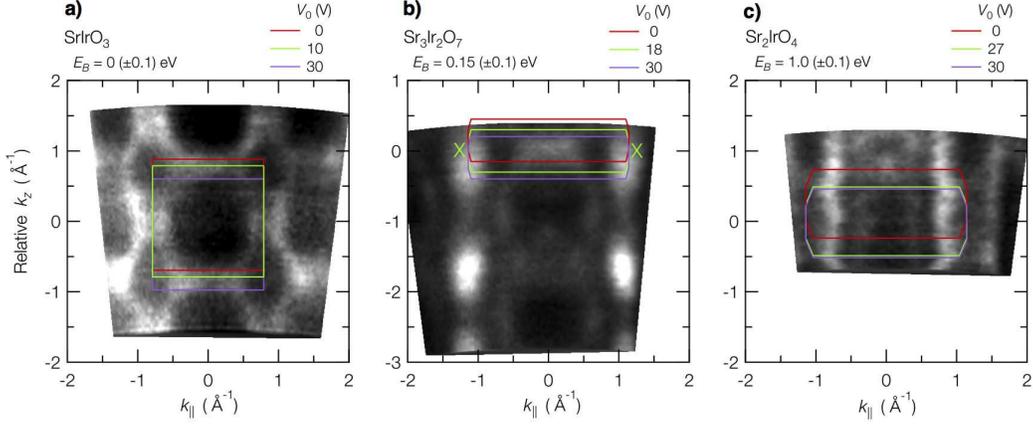}
\caption{
Constant-energy surfaces along the $k_z$ axis in (a) Sr113, (b) Sr327, and (c) Sr214.
The cross-sectional views of the BZs are  displayed by solid lines.
}
\label{Fig_S1}
\end{figure*}

As shown in Figs.~\ref{Fig_S1}(a)-(c),  a Brillouin zone (BZ)   shifts relative to the constant-energy surface depending on the inner potential $V_0$.
For SrIrO$_3$ (Sr113), it is obvious that the best-fit value of $V_0$ occurs at  about 10~eV within any realistic value  (0$\leqslant$ $V_0$ $\leqslant$ 30~eV).
In contrast, it is difficult  to optimize the $V_0$ for Sr$_3$Ir$_2$O$_7$ (Sr327) and Sr$_2$IrO$_4$ (Sr214)
 since   the electronic structures along the $k_z$ axis in Sr327 do not follow the periodicity of the BZ
because of the  $k_z$-dependent  oscillation resulting from the bilayer coupling, while in Sr214, they are almost structureless because of the layered crystal structure.
We  estimated  the $V_0$ of Sr327  to be 18~eV, where the electronic structure around a $k_z$ point, indicated as ``$\Gamma$''  shown in Fig.~\ref{Fig_S1}(b),
 is symmetric about the determined high-symmetry  line X-$\Gamma$-X.
As a consequence, the waveform of the bilayer coupled oscillation is well-reproduced by the calculation shown in Fig.~5(b).
In Sr214, we found a slight deformation in the constant-energy surface at $E_B$=1.0~eV, as shown in Fig.~\ref{Fig_S1}(c).
To fit the BZ to the deformed surface, a $V_0$ value of about 27~eV was used for Sr214.

\section{Determination of binding energies for producing constant-energy surfaces}

\begin{figure}
\includegraphics[width=6.0cm,clip]{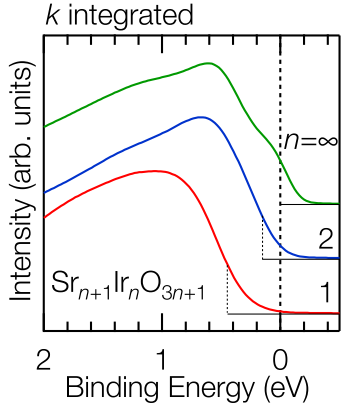}
\caption{
Referenced  photoemission spectra of RP-iridates.
}
\label{Fig_S2}
\end{figure}

Figure~\ref{Fig_S2} shows  the photoemission spectra  obtained by  summing up the  $k$-resolved spectra in both   $k_x$-$k_y$  and $k_{||}$-$k_z$ planes  shown in Figs.~1(d)-(i).
We  referred the spectra to produce the constant-energy surfaces  in Figs.~1(e), (f), (h), and (i).
The values of $E_B$  in these figures were chosen such that the spectral intensities of Sr327 or Sr214  at  $E_B$ 
was similar to that of Sr113 at the Fermi energy ($E_F$), as indicated by the dotted lines in Fig.~\ref{Fig_S2}.

\section{Estimation of Fermi velocity and Fermi wavenumber}

\begin{figure*}
\includegraphics[width=12cm,clip]{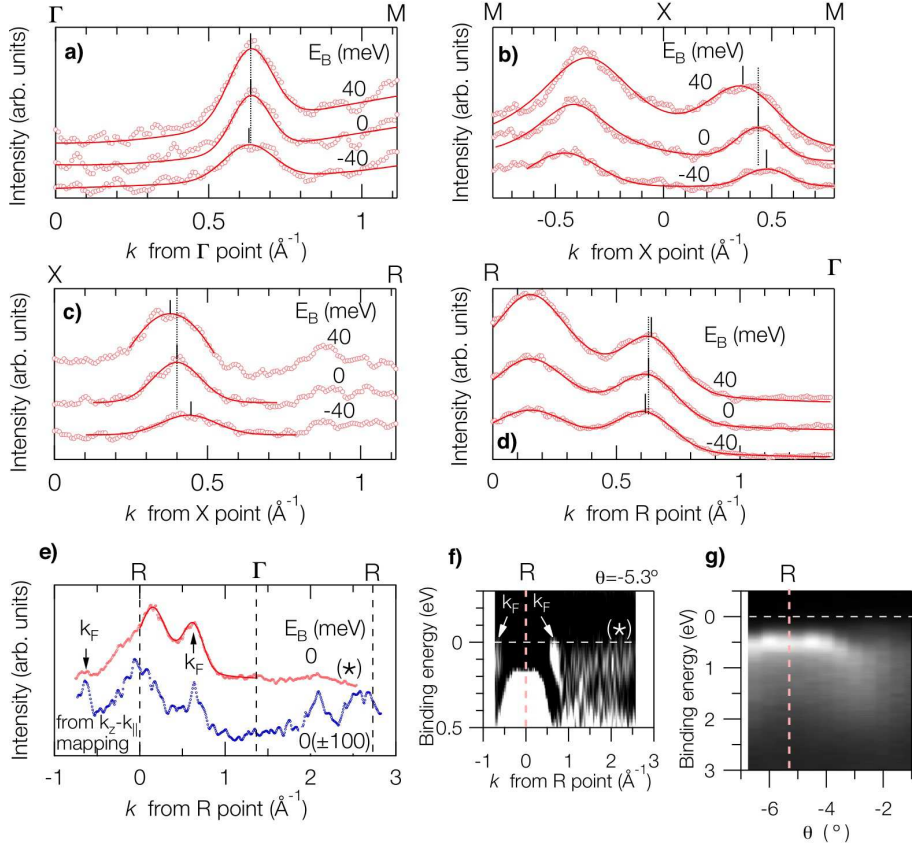}
\caption{
Estimation of  peak positions in momentum-distribution curves along the (a) $\Gamma$-M, (b) X-M, (c) X-R, and (d) R-$\Gamma$ lines in Sr113.
Only some examples of the curves in the vicinity of $E_F$ are shown.
To fit the experimental data, one or two  Gaussian  functions, and if necessary, a background Lorentzian function were used.
(e) Momentum-distribution curves at $E_F$ extracted from the ARPES spectrum partly shown in Fig.~3(f) and from the constant-energy surface in Fig.~1(d).
(f) The second-derivative image of the ARPES spectrum along the R-$\Gamma$ line.
The momentum-distribution curve denoted by (*) in (e) corresponds to the data indicated by the broken line at $E_F$. 
(g) Energy-band dispersion along the $\theta$ axis, which is perpendicular to the horizontal axis in (f).
}
\label{Fig_S3}
\end{figure*}

Figures~\ref{Fig_S3}(a)-(d) show  momentum-distribution curves of Sr113 in the vicinity of $E_F$.
We  used one (two) Gaussian function(s) to reproduce the curves along the $\Gamma$-M and X-R (M-X and R-$\Gamma$) lines,
in addition to a Lorentzian-type background  function if necessary.
 The estimated peak position indicated by the vertical bar significantly shifts with the binding energy in the X-M and X-R lines.
 Meanwhile, the shift is not well-resolved in the $\Gamma$-M and R-$\Gamma$ lines because of the large $v_F$ values.
 
The momentum-distribution curve at $E_F$ along the R-$\Gamma$ line is shown in Fig.~\ref{Fig_S3}(e) (in red); this was  obtained 
using the measurement with high energy resolution and high statistics.
The cross-sectional curve of the mapped  Fermi surface in the $k_z$-$k_{||}$ plane is  depicted in blue.
Compared to the blue line, the  intensity of the red line on the negative side of the $k$ axis is strongly suppressed due to the matrix element effect.
Nevertheless,  the R point should be correctly assigned  since the two $E_F$-crossing points and a parabolic band centered at the R point
are clearly resolved in Fig.~\ref{Fig_S3}(f).
In addition,  Fig.~\ref{Fig_S3}(g) shows that the band maximum of the parabolic band was also located at the assigned R point in the other axis perpendicular to the horizontal axis in Fig.~\ref{Fig_S3}(f), demonstrating a genuine high-symmetry point.

\section{Visible energy bands at different photon energies}

\begin{figure*}
\includegraphics[width=8.6cm,clip]{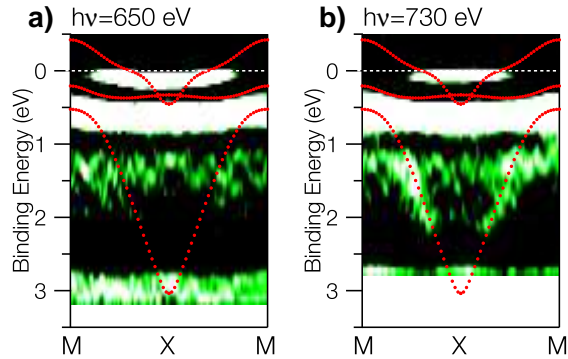}
\caption{
Energy-band dispersion along the X-M line in Sr113 measured at two different photon energies: (a) 650~eV and (b) 730~eV.
The second-derivative images  yield a better view of the dispersions.
The LDA+SOC band structure of  cubic Sr113 is also shown by dots.
}
\label{Fig_S4}
\end{figure*}

Figures~\ref{Fig_S4}(a) and (b) show  the energy-band dispersions along the X-M line in the Sr113 measured at two different photon energies.
One can see the widely dispersive band in  the $E_B$=1-2~eV range at $h\nu$=730~eV, which are not resolved at 650~eV.
In contrast, the characteristic feature in the vicinity of $E_F$ is more clearly seen at $h\nu$=650~eV.

\section{Deformation of Fermi surface}

As listed in Table~\ref{Table_S2}, we have obtained the Fermi wavenumber $k_F$ of Sr113  from the ARPES spectra and LDA+SOC calculations shown in Figs.~3(e) and (f).
Experimental $k_F$ values are generally closer to the values in  TB model  than to those in  LDA+SOC calculations.

\begin{table}
\caption{
Fermi wavenumber $k_F$ along high-symmetry lines in  Sr113.
$k_F$ is defined as the distance from the point M (R)  in the X-M and M-$\Gamma$ ($\Gamma$-R and R-X) lines.
The value in  square brackets represents the deviation from   $k_F$ in each high-symmetry line of the experimental FS listed at the top of the Table.
}
\begin{ruledtabular}
\begin{center}
\begin{tabular}{lcc}

{} &high-symmetry line & $k_F$  \\ 
{} & in cubic BZ & ({\AA$^{-1}$}) \\ \hline
\\
SX-ARPES & X -- M & 0.37\\
 & M -- $\Gamma$ & 0.48\\
 & $\Gamma$ -- R & 0.63\\
 & R -- X  & 0.71\\
 \\
LDA+SOC  & X -- M & 0.48 [130\%]\\
 & M -- $\Gamma$ & 0.32 [67 \%] \\
 & $\Gamma$ -- R & 0.56 [89 \%] \\
 & R -- X & 0.87 [122\%] \\
  \\
TB model  & X -- M & $\pi/2a$  [107\%]\\
 & M -- $\Gamma$ & $\sqrt{2}\pi/3a$  [78 \%]\\
 & $\Gamma$ -- R & $\sqrt{3}\pi/2a$   [108\%]\\
 & R -- X & $2\sqrt{2}\pi/3a$  [104\%]\\

\end{tabular}
\end{center}
\end{ruledtabular}
\label{Table_S2}
\end{table}


\begin{figure*}
\includegraphics[width=12.5cm,clip]{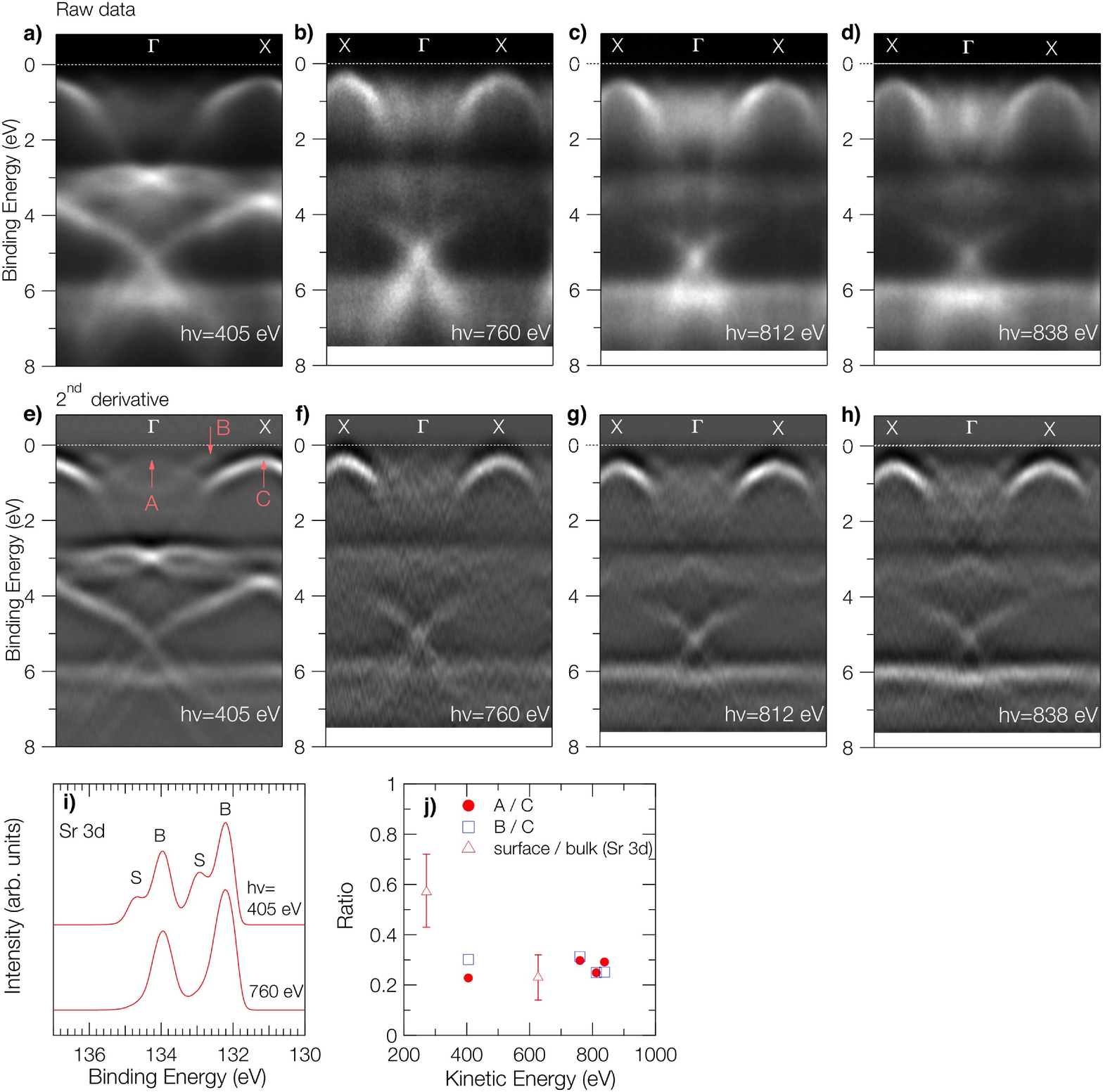}
\caption{
Energy-band dispersions of Sr214 measured at various photon energies:
(a)-(d) photoelectron-intensity distributions and (e)-(h) their second-derivative images.
(i) Sr 3$d$ core-level photoemission spectra at $h\nu$=405 and 760~eV.
``B'' and ``S'' denote the peaks originating from the electronic states in the bulk and on the surface, respectively.
(j) The  ratio of  the photoelectron intensity at point $\Gamma$  in a folded $j_{\rm eff}$=3/2 band [``A'' in (e)] to that at  point X in  the original unfolded band (``C'').
The ratio between the intensities  around ($\pi$/2,$\pi$/2,0) in the $j_{\rm eff}$=1/2 band (``B'') and at the X point in the  $j_{\rm eff}$=3/2 band is also shown.
}
\label{Fig_S5}
\end{figure*}

\section{${\rm \emph{h}}$$\boldsymbol{\nu}$-dependence of photoelectron intensity in folded and unfolded ${\rm \emph{j}}$$_{{\bf{eff}}}$ bands}

Figures~\ref{Fig_S5}(a)-(h) show the energy-band dispersions along the $\Gamma$-X line in the Sr214 measured at various photon energies.
To investigate the influence of the photon-energy variation on the photoelectron intensity in some $j_{\rm eff}$ bands,
we  focused on three points named A, B, and C, indicated by arrows in Fig.~\ref{Fig_S5}(e).
The bands to which  points A and C belong are assigned to the folded  and unfolded $j_{\rm eff}$=3/2 bands, respectively, 
predicted by  calculations based on the three-orbital Hubbard model. 
The band including  point B  is predicted in both the folded and unfolded BZ pictures.
As  seen in Fig.~\ref{Fig_S5}(j),  photoelectron intensities at  points A and B relative to  point C are independent of the kinetic energy of the photoelectrons.
The surface spectral weight relative to the bulk weight estimated from the Sr $3d$ core-level photoemission spectrum in Fig.~\ref{Fig_S5}(i) is also shown, clearly indicating that 
the surface components are suppressed with increasing  photon energy, because of the variation in bulk sensitivity.


\begin{references}
\bibitem{Bulk_sensitivity}
The photoelectron escape depth $\lambda$ in the RP-iridates was found to be 
about 14~{\AA} at h$\nu$$\simeq$750~eV using the TPP-2M formula~\cite{Tanuma94}.
The  $\Delta k_z$ was then roughly estimated from $\lambda\cdot\Delta k_z\approx1/2$ based on the Heisenberg uncertainty relation.
\bibitem{note4}
The $k$-mesh of (10$\times$10$\times$10) in the first BZ and the muffin-tin radii of 2.23, 1.97, 1.74 bohr for Sr, Ir, and O were used,
being the same as in Ref.~[\onlinecite{Arita12}].
Overall feature in our calculated dispersion relation is quite similar to the reported one in Ref.~[\onlinecite{Nie15}].
\bibitem{note2}
Judging from our calculated results, the band originates from the $j_{\rm eff}$=3/2 ($j_z$=$\pm$1/2) band although it was assigned to the $j_{\rm eff}$=1/2 band.
\bibitem{note3}
For instance, one can find only one band at $\Gamma$ point in the result at h$\nu$=85~eV (see Table~\ref{Table_2}) since the dispersion along ($\pi$,$\pi$)-($\pi$,0) is probably 
displayed  as that along the $\Gamma$-M line.
\bibitem{note5}
We measured  the ARPES spectra  above $T_N$ both at the end of the heating process  and at the start of the cooling process after cleaving the crystal.
 These spectra showed good reproducibility.
\end{references}
\end{document}